\documentclass[a4paper,fleqn,usenatbib]{mnras}

\usepackage{Times}

\usepackage[T1]{fontenc}
\usepackage{ae,aecompl}

\usepackage{graphicx}	
\usepackage{amsmath}	
\usepackage{amssymb}	

\title[The $\alpha$-enhancement relation in SFGs]{Star-forming galaxies are predicted to lie on a fundamental plane of mass, star formation rate and $\alpha$-enhancement}

\author[Matthee \& Schaye]{Jorryt Matthee$^{1}$\thanks{E-mail: matthee@strw.leidenuniv.nl} \& Joop Schaye$^{1}$\\
$^{1}$ Leiden Observatory, Leiden University, P.O.\ Box 9513, NL-2300 RA Leiden, The Netherlands\\}

\date{Accepted 2018 May 17. Received 2018 May 17; in original form 2018 February 19.}

\pubyear{2018}

\begin{document}
\label{firstpage}
\pagerange{\pageref{firstpage}--\pageref{lastpage}}
\maketitle

\begin{abstract}
Observations show that star-forming galaxies reside on a tight three-dimensional plane between mass, gas-phase metallicity and star formation rate (SFR), which can be explained by the interplay between metal-poor gas inflows, SFR and outflows. However, different metals are released on different time-scales, which may affect the slope of this relation. Here, we use central, star-forming galaxies with M$_{\rm star}=10^{9.0-10.5}$ M$_{\odot}$ from the EAGLE hydrodynamical simulation to examine three-dimensional relations between mass, SFR and chemical enrichment using absolute and relative C, N, O and Fe abundances. We show that the scatter is smaller when gas-phase $\alpha$-enhancement is used rather than metallicity. A similar plane also exists for stellar $\alpha$-enhancement, implying that present-day specific SFRs are correlated with long time-scale star formation histories. Between $z=0$ and 1, the $\alpha$-enhancement plane is even more insensitive to redshift than the plane using metallicity. However, it evolves at $z>1$ due to lagging iron yields. At fixed mass, galaxies with higher SFRs have star formation histories shifted toward late times, are more $\alpha$-enhanced and this $\alpha$-enhancement increases with redshift as observed. These findings suggest that relations between physical properties inferred from observations may be affected by systematic variations in $\alpha$-enhancements.
\end{abstract}




\section{Introduction}
One of the key goals of galaxy formation studies is to understand when and how the interstellar medium (ISM) was chemically enriched and how the metal abundances of stars evolved through cosmic time. A key observable is the metallicity of the ISM (Z$_{\rm gas}$), which controls the metallicity of the stars that are being formed. Gas-phase metallicity is sensitive to gas accretion (pristine or recycled), chemical enrichment through stellar mass loss and supernovae, and outflows driven by feedback associated with the formation of stars and the growth of super-massive black holes.

Observationally, Z$_{\rm gas}$ is typically quantified by the (light-weighted) oxygen to hydrogen abundance in H{\sc ii} regions, (O/H). Can the choice for this ratio be theoretically motivated? In principle it can, as oxygen is the most abundant metal in the Universe and relatively easily observable. 

However, in a gas-reservoir that evolves due to inflows and outflows, changes in (O/H)$_{\rm gas}$ may simply reflect variations in the gas fraction. Indeed, variations in gas-phase metallicities at fixed galaxy mass are observed to be anti-correlated with variations in specific star formation rates (sSFRs; e.g. \citealt{Ellison2008}), a galaxy property that is closely related to the gas fraction \citep[e.g.][]{Bothwell2013}. As a result, star-forming galaxies in the present-day Universe reside on a three dimensional relation between M$_{\rm star}$, SFR and Z$_{\rm gas}$ \citep[e.g.][]{LaraLopez2010,Mannucci2010}, called the fundamental metallicity relation (FMR; \citealt{Mannucci2010}). This correlation weakens/breaks down at masses M$_{\rm star}>10^{10.5}$ M$_{\odot}$ \citep[e.g.][]{Yates2012,Zahid2013,Salim2014}, potentially due to the increased importance of AGN feedback \citep{DeRossi2017}. 

The existence of a 3D metallicity relation is a well-understood property of galaxy equilibrium models \citep[e.g.][]{FinlatorDave2008,Dave2012,Lilly2013,RP2016} and is also reproduced by cosmological hydrodynamical simulations \citep[e.g.][]{Lagos2016,DeRossi2017}. An increase in the accretion of (metal poor) gas both lowers the metallicity and fuels an increase in the SFR, as long as the time-scales on which the SFR and Z$_{\rm gas}$ evolve, are of the same order \citep{Torrey2017}. This means that the 3D metallicity relation reflects the time-scales of variations in the gas mass fraction \citep[e.g.][]{Bothwell2013,Forbes2014,Lagos2016,Brown2017}. 

What happens if we use another metal as the tracer of metallicity? Different metal species are produced on varying time-scales \citep[e.g.][]{Tinsley1979}. The enrichment time-scale for oxygen is short ($\sim 10$ Myr) as it is only produced in massive stars. Significant contributions ($\approx 30$ \%) from AGB stars result in longer time-scales for carbon and nitrogen ($\sim200$ and 500 Myr, respectively; see \citealt{Wiersma2009enrich}). Significant iron enrichment occurs over the longest time-scale ($\gtrsim2$ Gyr) as a substantial fraction (20-30 \%) is produced in Type Ia supernovae (SNe). As a result, the different time-scales on which star formation histories (SFHs) and gas fractions change, drive variations in relative metal abundances.

In this letter, we use the EAGLE cosmological simulation to investigate which chemical abundance is most closely related to stellar mass and SFR and which scaling relation evolves least between $z=0-2$. This letter is structured as follows. We first discuss the simulation and the various definitions of metallicity in \S $\ref{sec:methods}$. \S $\ref{sec:howmuchscatter}$ reports how much scatter there is in the 2-(3-)dimensional mass - metallicity (- SFR) relation for different element abundances and how this evolves with cosmic time. In \S $\ref{sec:observational_implications}$ we discuss the implications of our results for the interpretation of observations and we discuss the prospects for testing our results observationally.

\section{Methods} \label{sec:methods}
\subsection{The EAGLE simulations}
We use galaxies from the (100 cMpc)$^3$ reference model\footnote{We have verified that our conclusions remain unchanged when we use the higher-resolution (25 cMpc)$^3$ RECAL model of the EAGLE simulation. The mass-metallicity relation is  steeper in the RECAL model (see also \citealt{Schaye2014}), the normalisation of abundances are lower (at fixed mass) and $\alpha$-enhancement is higher by 0.1 dex. However, the slopes of the correlations between sSFR and chemical abundances at fixed mass remain unchanged.} from the EAGLE cosmological hydrodynamical simulation \citep{Schaye2014,Crain2015,McAlpine2015}. EAGLE uses a modified version of the smoothed particle hydrodynamics (SPH) code {\sc Gadget3} \citep{Springel2005Gadget} with a modern formulation of SPH (see \citealt{SchallerSPH}).

Gas particles with sufficiently high density (depending on metallicity) are transformed into star particles following the pressure-dependent implementation of the Kennicutt-Schmidt law of \cite{SchayeVecchia2008}. Each star particle is a single stellar population with a \cite{Chabrier2003} initial mass function and the metallicity inherited from the parent gas particle. As detailed in \cite{Wiersma2009enrich}, the simulation tracks the enrichment of eleven elements as they are released into the ISM on their relevant time-scales. This model includes yields from type II SNe, mass loss from intermediate-mass AGB stars and winds from high mass stars and type Ia SNe. The cosmic type Ia SN rate follows an exponential distribution, where the normalisation and e-folding time-scale of 2 Gyr are chosen to reproduce the evolution of observed SN Ia rates (see \citealt{Schaye2014}).\footnote{We note that the delay time distribution of SNIa in EAGLE is somewhat shifted to longer times compared to recent observations \citep[e.g.][]{Maoz2012,Friedmann2018}. As a result, $\alpha$-enhancements in EAGLE may be somewhat over-estimated, though uncertainties in nucleosynthetic yields may be more important \citep[e.g.][]{Wiersma2009enrich}.} The radiative cooling rates are computed on an element-by-element basis \citep{Wiersma2009cooling}. Energy feedback from star formation is implemented stochastically \citep{VecchiaSchaye2012} and was calibrated to reproduce the present-day galaxy stellar mass function and sizes. The growth of black holes and the feedback associated with AGN activity are also modelled \citep{Springel2005BH,Rosas2015,Schaye2014}. The simulation uses baryonic particles with (initial) mass $1.8\times10^6$ M$_{\odot}$ and a gravitational force resolution better than 0.7 physical kpc. 

Our analysis includes central galaxies\footnote{To avoid strong environmental effects, we focus on central galaxies. See \cite{Bahe2017} for a study of the differences in the metallicities of central and satellite galaxies in EAGLE.} with M$_{\rm star} = 10^{9-10.5}$ M$_{\odot}$ at all redshifts and we select star-forming galaxies with sSFR $>10^{(-11, -10.4, -10)}$ yr$^{-1}$ at $z=(0.1, 1.0, 2.0)$, respectively. These cuts encompass the main sequence of star-forming galaxies as defined by \cite{Elbaz2007}. They correspond to galaxies that have at least $\approx$ 500 star particles and $\approx 30$ star-forming gas particles and ensure a sufficient number of galaxies (6201, 7046, 5487 at $z=0.1, 1.0, 2.0$, respectively). The upper mass limit is chosen to ensure that the star formation histories are not yet significantly influenced by AGN feedback, which weakens 3D chemical enrichment correlations \citep{DeRossi2017}.
 
\begin{figure*}
\begin{tabular}{cc}
\includegraphics[width=8.55cm]{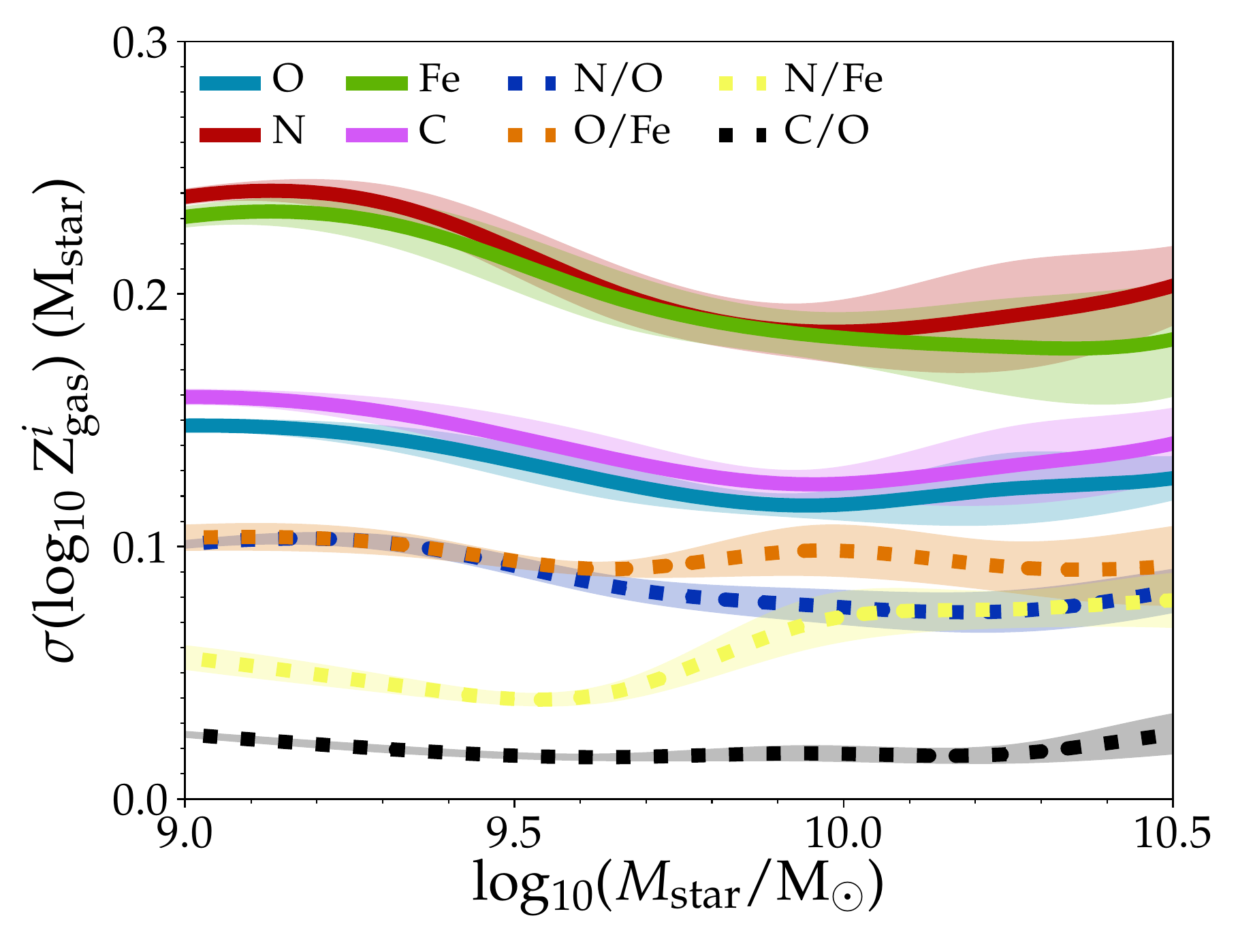}&
\includegraphics[width=8.55cm]{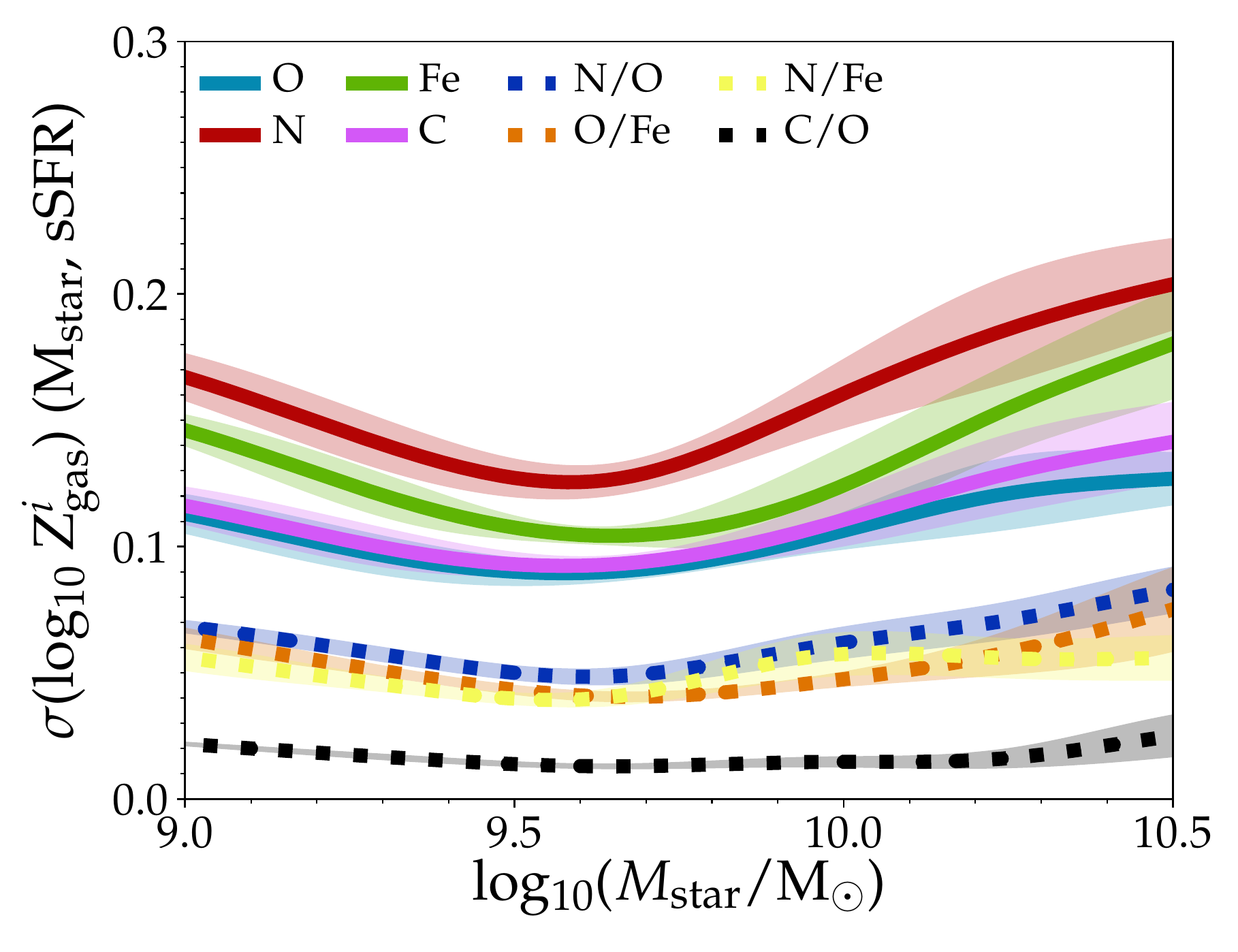}\\
\end{tabular}
\caption{\small{The scatter in the mass- gas metallicity relations for central star-forming galaxies at $z=0.1$, as a function of stellar mass. The {\it left} panel shows the scatter in the 2D relation, while the {\it right} panel shows the scatter around the 3D relation (i.e. including log$_{10}$ sSFR). Different colours show different metal abundances. Solid lines show absolute abundances, while dashed lines show relative abundances. All abundances are for the star-forming gas-phase. }}
\label{fig:scatter_Zgas_z0p1}
\end{figure*}
 
\subsection{Definitions \& choice of metal species}
We use SPH-smoothed metallicities in stars and in the star-forming gas-phase (see \citealt{Schaye2014} for details). Metal abundances are expressed relative to the Solar values.\footnote{We adopt a solar metallicity Z$_{\odot} = 0.012947$. We follow the standard notation of [O/Fe] = (O/Fe) - (O/Fe)$_{\odot}$. We adopt solar abundances of (O/Fe)$_{\odot}$ = log$_{10}(X^{\rm O}_{\odot}/{X^{\rm Fe}_{\odot}}) = 0.647$ and (N/O$)_{\odot}$ = log$_{10}({X^{\rm N}_{\odot}}/{X^{\rm O}_{\odot}})$ = -0.92 \citep{Asplund2009}, where $X^{i}$ is the mass-fraction of element $i$.} We investigate the metals C, N, O and Fe, because of their abundance, observability and importance for determining the properties of stars. 

Besides investigating the mass fractions of these metals, we also investigate their relative abundances. In the EAGLE model, the C/O ratio of stellar ejecta is sensitive to systematic changes in SFRs on $\sim200$ Myr time-scales. Nitrogen is produced on longer time-scales than C and O and is a secondary element (increased stellar nitrogen abundance increases the nitrogen yields). Therefore, N/O is a measure of the integrated chemical evolution history. The O/Fe ratio ($\approx\alpha$-enhancement) traces even longer time-scales and is therefore sensitive to the entire SFH of a galaxy. As both nitrogen and iron trace relatively long time-scales, variations in the N/Fe ratio are sensitive to differences at early times that are amplified by the secondary nature of nitrogen. We do not explicitly discuss C/Fe as we find that this ratio is similar to O/Fe.

\section{The scatter in 2D and 3D mass - metallicity relations} \label{sec:howmuchscatter}
We first focus in \S $\ref{sec:2D}$ on 2D relations between mass and gas-phase metal abundances in the late Universe ($z=0.1$), then in \S $\ref{sec:3D}$ measure the scatter in 3D relations that account for variations in sSFR and we finish in \S $\ref{sec:evolution}$ with investigating evolution. We measure scatter as the standard deviation of log$_{10}$(Z$^i$) in rolling bins of stellar mass (with bin-width 0.3 dex and with bin-centers in increasing steps of 0.1 dex) and interpolate these values to obtain a smoothed trend. Here, Z$^i$ is either the metal mass-fraction or the relative metal abundance.

We account for variations in log$_{10}$ sSFR by fitting a linear relation ${\rm log}_{10} {\rm Z}^i = a + b\times {\rm log}_{10} {\rm sSFR}$ in each mass bin and measuring the scatter in ${\rm log}_{10} {\rm Z}^i_{\rm 3D} = {\rm log}_{10} {\rm Z}^i - [a(M_{\rm star}) + b(M_{\rm star})\times {\rm log}_{10} {\rm sSFR}]$. We perform our measurements in eight separate sub-volumes of (50 cMpc)$^3$ and show the median and its standard deviation to highlight variations due to cosmic variance. The results are shown in Figure $\ref{fig:scatter_Zgas_z0p1}$, where we plot $\sigma$(log$_{10}$ Z$^i$) as a function of stellar mass.

\subsection{Absolute metal abundances} \label{sec:2D}
The left panel of Fig. $\ref{fig:scatter_Zgas_z0p1}$ shows that among ISM metal abundances, there is least scatter in the oxygen mass fraction ($\approx 0.15$ dex, similar to observations of O/H abundances; \citealt{Tremonti2004}), followed closely by the C mass-fraction. There is significantly more scatter in the N and Fe fractions.\footnote{We find qualitatively similar results for mass-weighted stellar abundances. The typical scatter in stellar abundances and the differences between different elements are smaller, but the rank-order is preserved.}  

The right panel of Fig. $\ref{fig:scatter_Zgas_z0p1}$ shows the scatter in metallicities as a function of mass after accounting for the trend with sSFR. The scatter is reduced for all absolute abundances (i.e. mass fractions), but we find that it is most strongly reduced for Fe. This means that at fixed mass, fluctuations in sSFR are most closely related to the abundance of iron, and least strongly related to that of oxygen. However, the scatter in the 3D metallicity relation is still smallest for oxygen.

\begin{figure}
\includegraphics[width=8.5cm]{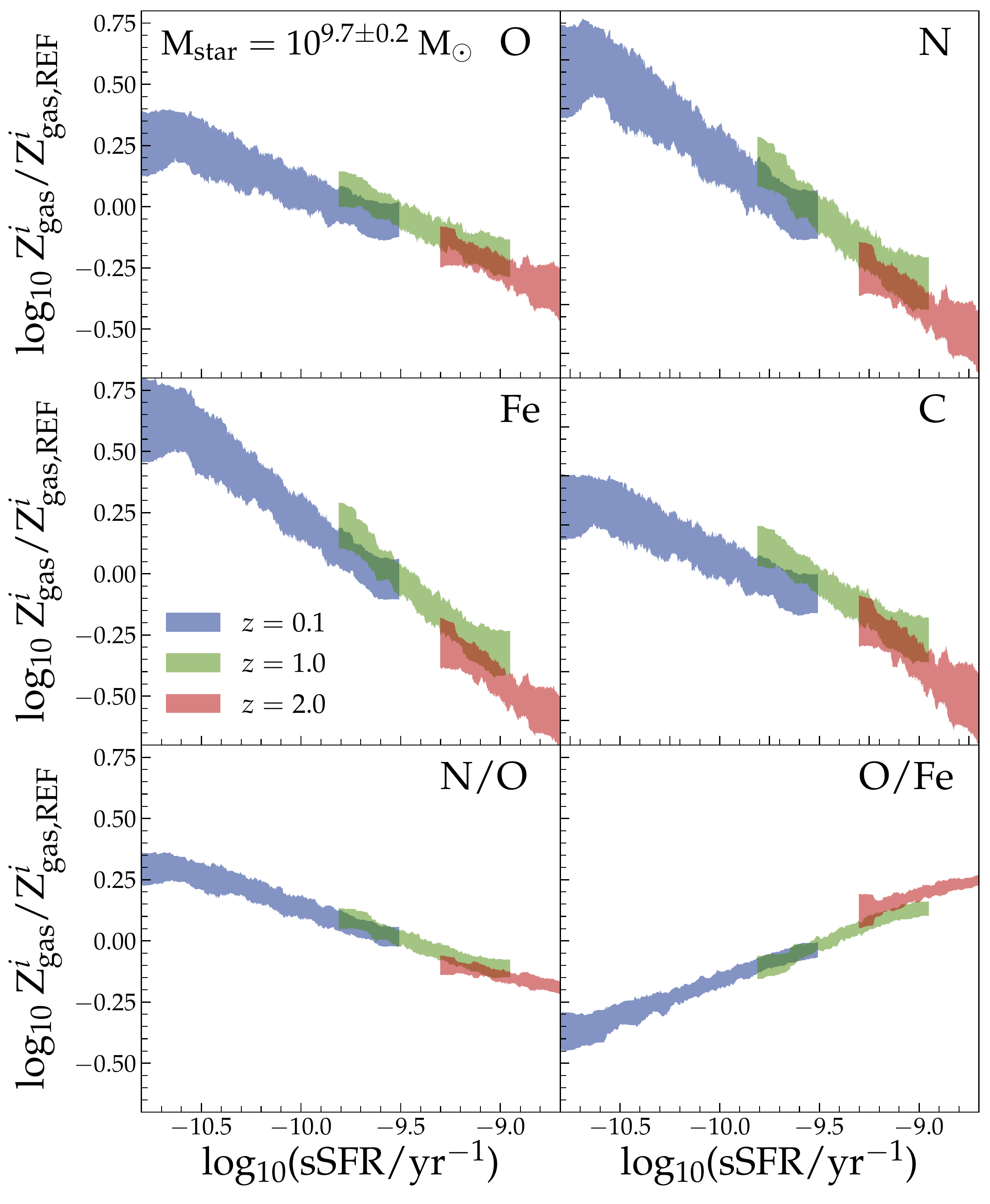}\\
\caption{\small{Evolution of the relations between the chemical enrichment of the ISM and sSFR for central star-forming galaxies with M$_{\rm star}=10^{9.7\pm0.2}$ M$_{\odot}$. Each panel shows a different abundance ratio that is scaled to a {\it reference} median abundance at log$_{10}$(sSFR$_{z=1.0}$/yr$^{-1}$) $= -9.5$ to facilitate the comparison. Absolute metal abundances show a weak evolution in the normalisation of the relation, particularly between $z=0.1$ and 1.0. The N/O ratio correlates  more weakly with sSFR, but the correlation evolves less. Fe/O does not evolve between $z=0.1$ and 1.0, but shows an offset at $z=2.0$ because the formation time-scale of iron is no longer small compared with the age of the Universe.}}\label{fig:FMR_evo}
\end{figure}

The differences in scatter shown in the left panel of Fig. $\ref{fig:scatter_Zgas_z0p1}$ must arise from variations in the N/O and O/Fe ratios and are therefore a consequence of the different enrichment time-scales. If the gas fraction and hence sSFR increase, oxygen enrichment occurs almost simultaneously. This means that any increase in M$_{\rm H}$, decreasing the O/H ratio, is rapidly followed by an increase in M$_{\rm O}$ and hence O/H. This processes limits the scatter in the O/H - M$_{\rm star}$ relation. As the enrichment time-scales of N and Fe are significantly longer, this scatter-reducing process is less effective for the N and Fe abundances, resulting in a larger scatter. This picture is supported by a comparison of the two panels of Fig. $\ref{fig:FMR_evo}$, which shows that at fixed mass, variations in N and Fe mass fractions are more sensitive to fluctuations in sSFRs than is the case for O.

\subsection{Relative metal abundances} \label{sec:3D}
Relative metal abundances encode information about the chemical enrichment history and are less sensitive to present-day fluctuations in gas fractions. As a consequence, we find that N/O, C/O, N/Fe and O/Fe are more tightly correlated with M$_{\rm star}$ than any of the absolute abundances are (left panel of Fig. $\ref{fig:scatter_Zgas_z0p1}$, consistent with observations from e.g. \citealt{AndrewsMartini2013}). There is very little variation in C/O and N/Fe at fixed mass. There is more variation in abundance ratios that trace both short and long time-scales (N/O and O/Fe), which is a result of galaxies at fixed stellar mass having different SFHs.

The scatter in the N/O and O/Fe abundance ratios decreases once fluctuations in sSFR are accounted for. This indicates that present-day sSFR is related to long-time-scale differences in chemical enrichment histories (and hence SFHs). This can also be seen in Fig. $\ref{fig:evolution_alpha}$, which shows that the mass-weighted stellar O/Fe correlates with sSFR. Because stellar O/Fe decreases with each cycle of star formation, this implies an anti-correlation between sSFR and the SFH. Moreover, at fixed mass, gas-phase O/Fe is strongly correlated with stellar O/Fe (with a Spearman rank correlation R$_S=0.7$) and anti-correlated with mass-weighted stellar age (R$_S=-0.75$). The origin of the correlation between present-day sSFR and SFH lies in the fact that, at fixed mass, sSFR anti-correlates with the dark matter halo formation redshift. This manifestation of galaxy assembly bias is explored in more detail in \cite{MattheeSchaye2018b}.

\begin{figure}
\includegraphics[width=8.5cm]{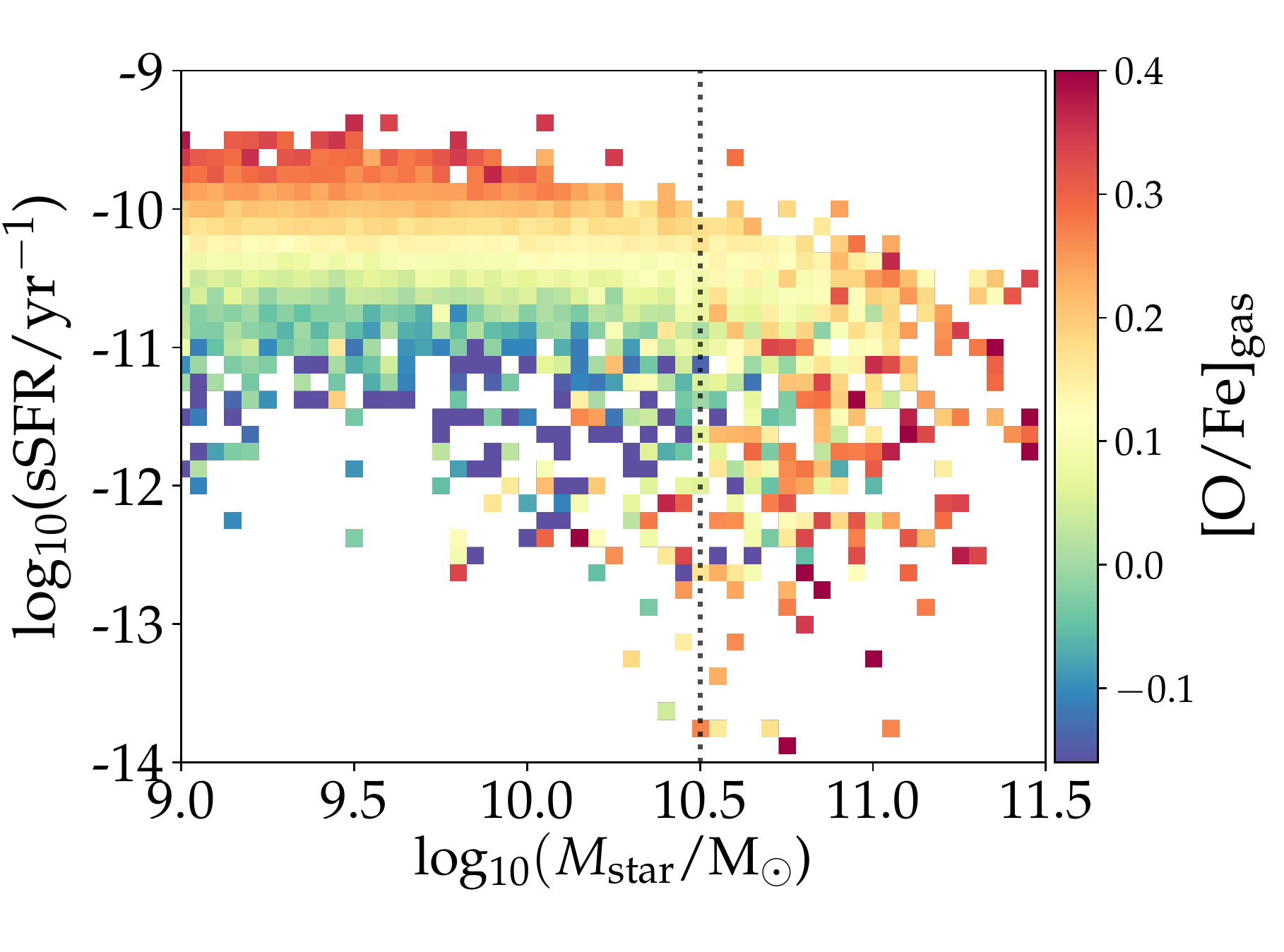}\\
\caption{\small The correlation between the scatter in sSFR(M$_{\rm star}$) and [O/Fe] clearly illustrates the 3D relation between M$_{\rm star}$ - sSFR - O/Fe, for M$_{\rm star}<10^{10.5}$ M$_{\odot}$. At higher masses the 3D correlation weakens and $\alpha$-enhancement is mostly set by M$_{\rm star}$. }\label{fig:SFRMstar_alpha}
\end{figure}

Moreover, our results mean that the 3D relations M$_{\rm star}$ - sSFR - O/Fe or N/O have less scatter than the `fundamental metallicity relation' with O/H. Although the scatter is even smaller if we use C/O or N/Fe, the residuals of their relations with mass do not correlate with sSFR. Hence, the small scatter merely reflects a similar nucleosynthetic origin and does not provide additional insight into galaxy evolution. We will therefore not consider these abundance ratios any further.

The reason for the tightness of the 3D $\alpha$-enhancement relation is that O/Fe combines two physical effects that work in the same direction (unlike O/H, where these effects act oppositely). The first is rapid O enrichment in response to a recent increase in SFR. The second is the lower Fe abundance for galaxies with high SFRs due to such galaxies tending to have delayed SFHs. Both effects give rise to an increasing $\alpha$-enhancement as a function of sSFR (this is also shown in Fig. $\ref{fig:FMR_evo}$).

We show a 2D projection of the M$_{\rm star}$ - sSFR - [O/Fe] relation in Fig. $\ref{fig:SFRMstar_alpha}$, where the third dimension of the relation, [O/Fe], is illustrated with the color-coding. For M$_{\rm star}<10^{10.5}$ M$_{\odot}$, the scatter in the M$_{\rm star}$ - sSFR relation is correlated strongly with [O/Fe], while the relation weakens at higher masses. Fitting a 3D plane to these data-points at $z=0.1$ using a least squares algorithm yields:  {$\rm [O/Fe]=-0.282\,log_{10}(M_{\rm star}/10^{10} M_{\odot}) +0.290\,log_{10}(SFR/M_{\odot} yr^{-1}) + 0.163$}. 

\begin{figure*}
\begin{tabular}{cc}
\includegraphics[width=8.6cm]{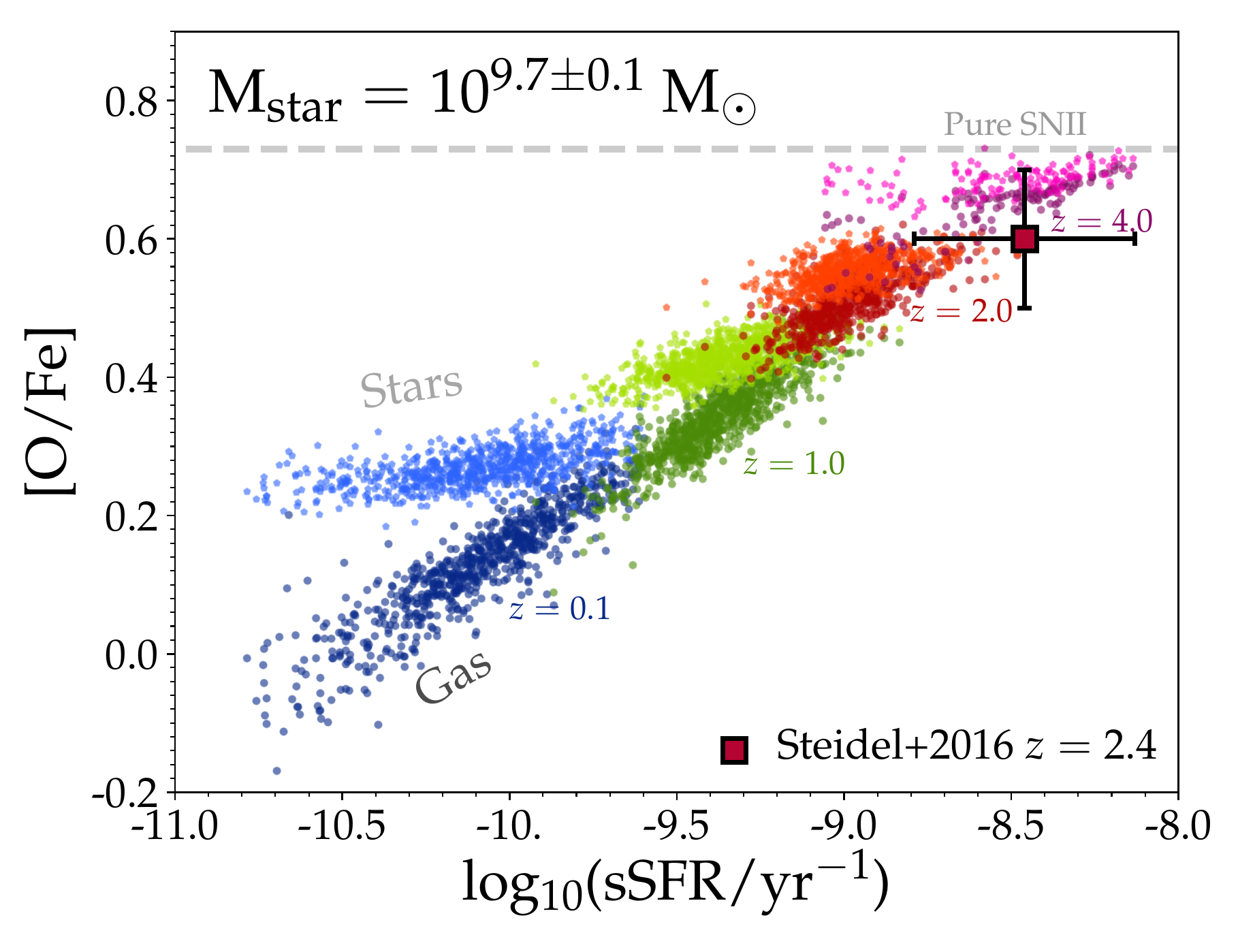}&
\includegraphics[width=8.6cm]{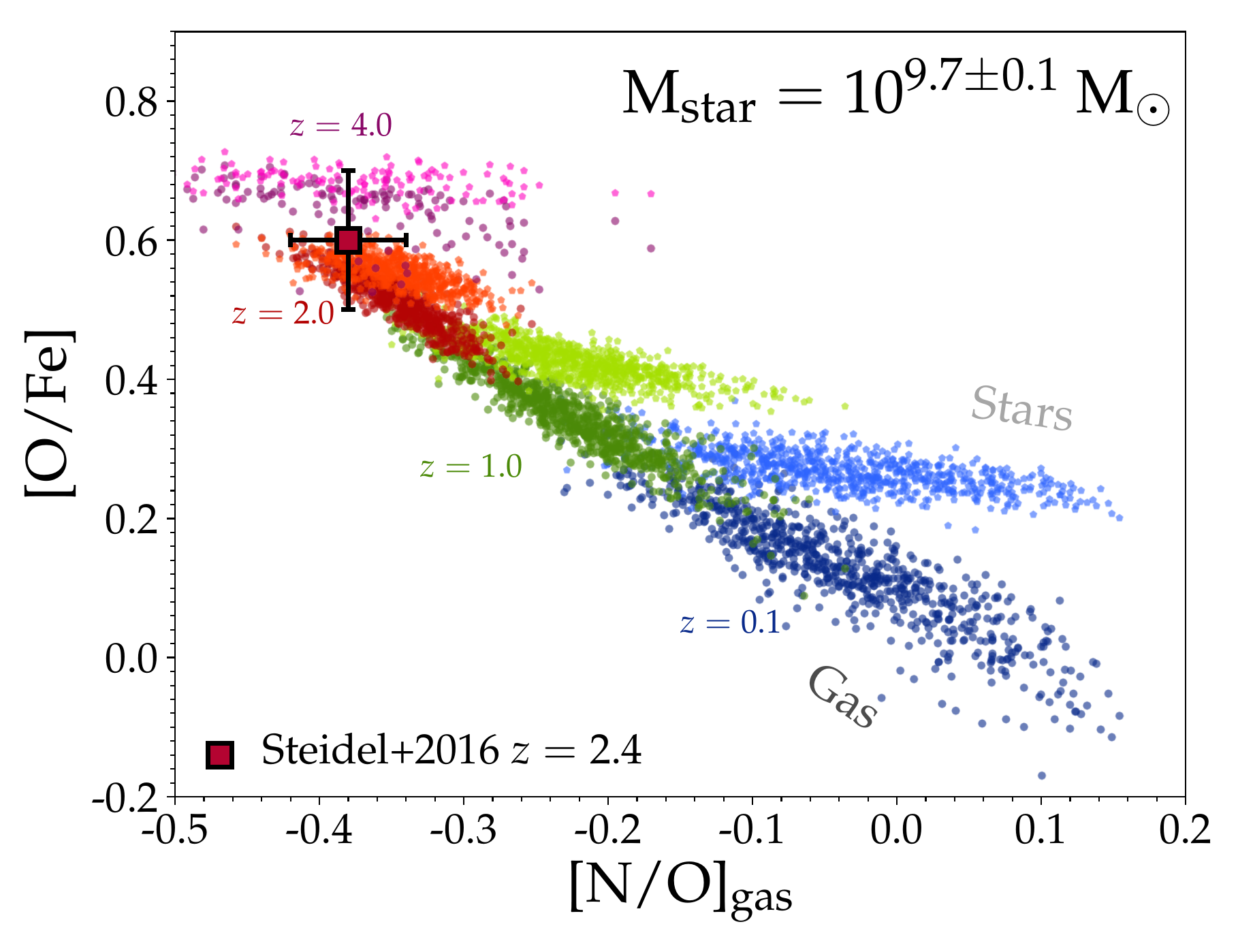}\\
\end{tabular}
\caption{\small{The relation between mass-weighted gas-phase and stellar $\alpha$-enhancement versus specific SFR ({\it left}) and versus gas N/O ratio ({\it right}) for different redshifts (different colours) for central star-forming galaxies with M$_{\rm star} = 10^{9.7\pm0.1}$ M$_{\odot}$. Light and dark colours correspond to stars and star-forming gas, respectively. At all redshifts, galaxies with a higher sSFR at fixed mass tend to be more $\alpha$-enhanced. At higher redshift, when SFHs are necessarily more compressed, the differences between gas and stellar abundances become smaller, typical $\alpha$-enhancements increase (similar to observational results from \citealt{Steidel2016}; data point) and the $\alpha$-enhancement converges towards the theoretical yields for massive stars \citep[dashed line;][]{Nomoto2006}. }}
\label{fig:evolution_alpha}
\end{figure*}

\subsection{Evolution through cosmic time} \label{sec:evolution}
The results described above are for $z=0.1$. What about higher redshifts? A key feature of the fundamental metallicity relation proposed by \cite{Mannucci2010}, is that it seems to be redshift-invariant \citep[see extended observational discussion in e.g.][]{Stott2013,Maier2014,Sanders2015,Sanders2017}. \cite{DeRossi2017} show that there is no evolution in the gas fraction - metallicity relation between $z=0-3$ in the EAGLE RECAL-L025N0752 simulation. To what extent does the same hold for the 3D $\alpha$-enhancement relation or relations using other metal abundances? 

Fig. $\ref{fig:FMR_evo}$ shows the correlation between sSFR and metal abundance (using different elements) at a fixed stellar mass of M$_{\rm star}=10^{9.7\pm0.2}$ M$_{\odot}$ and at $z=0.1, 1.0$ and $z=2.0$. We find that all absolute abundances anti-correlate with sSFR at all redshifts, but that the trends are slightly different for different elements (note that many trends identified in the previous subsection are clearly visible in this figure). To first order the different redshifts follow a single relation, but looking more closely, there is a slight shift in the normalisation between $z=0$ and 1 for all absolute abundances. Such a shift is not visible for N/O and O/Fe, indicating that the differences are due to differences in the H gas masses at $z=1$ compared to $z=0$. Indeed, at $z = 1$ the galaxies are much smaller than at $z = 0.1$ \citep{Furlong2017}, which, combined with the non-linear star formation law, implies that less gas is required to obtain the same SFR. This can explain the higher metallicity (at $z=1.0$ compared to $z=0.1$) at fixed sSFR.

The normalisation and slope of the relation between O/Fe  and sSFR do not evolve between $z=0$ and $z=1$, but the normalisation does shift slightly upwards at $z=2$. This is because at high redshift the formation of iron is still lagging (the $\alpha$-enhancement converges to the value for pure SNII yields at early times, see also Fig. $\ref{fig:evolution_alpha}$). There is a similar enrichment-lag for N/O that becomes more prominent at $z>3$ (not shown). Hence, at least for $z\lesssim1$ the relation between mass, SFR and $\alpha$-enhancement evolves even less than the ``fundamental metallicity relation''.

\section{Discussion} \label{sec:observational_implications}
\subsection{Observational implications}
In observations, Z$_{\rm gas}$ is typically measured using the relative strengths of strong emission lines from oxygen, nitrogen and hydrogen and calibrations based on local galaxies \citep[e.g.][]{PettiniPagel2004} or theoretical models. Besides metallicity, these line strengths implicitly depend on the ISM properties (density, temperature) and the ionising radiation, which depends on stellar age and iron abundance, among other properties \citep[e.g.][]{Eldridge2017}. It is therefore important to understand how these properties differ among galaxies and how they evolve with cosmic time.

As the left panel of Fig. $\ref{fig:evolution_alpha}$ shows, the gas-phase $\alpha$-enhancement is a strong function of sSFR. As a result, galaxies at higher redshift are more $\alpha$-enhanced. Such an increase has been inferred from observations (\citealt{Steidel2016}; data point in Fig. $\ref{fig:evolution_alpha}$), in quantitative agreement with the EAGLE prediction. 
Galaxies that are $\alpha$-enhanced have a lower iron abundance at fixed O/H, and hence have harder stellar spectra (as the hardness increases with lower iron abundance; see e.g. Figure 6 from \citealt{Eldridge2017}). The observed relative strengths of strong emission lines (such as H$\alpha$, [OIII] and [OII]) depend on the relative number of photons with ionisation energies corresponding to different line transitions. Therefore, variations in the hardness of the underlying ionisation field will result in different line ratios, even though oxygen abundances may be constant. Strong emission-line diagnostics may have to be revised, particularly when comparing samples with different redshifts.

Fig. $\ref{fig:evolution_alpha}$ shows that the mass-weighted stellar $\alpha$-enhancement of star-forming galaxies\footnote{Observations are light-weighted rather than mass-weighted. Light-weighted stellar abundances likely lie closer to the gas abundances. At high redshift the light- and mass-weighted stellar abundances converge due to the high sSFRs.} is also correlated with sSFR. As discussed above, this is strong evidence for a correlation between the present-day sSFR and the SFH. This correlation may also lead to biases, because the stellar light predicted by population synthesis models depends (or should depend) on $\alpha$-enhancement.

Fig. $\ref{fig:evolution_alpha}$ also shows that biases are not only expected between galaxy samples at different redshifts, but also at a fixed cosmic time. Differences in the $\alpha$-enhancements of galaxies add a systematic uncertainty correlated with the sSFR, which needs to be taken account in the applicability of emission line diagnostics.

\subsection{Prospects}
Can these predictions from the EAGLE simulation be tested observationally? Measuring the gas-phase $\alpha$-enhancement in star-forming galaxies is challenging due to the faintness of iron emission lines and uncertainties in dust-depletion corrections. However, we show in the right panel of Fig. $\ref{fig:evolution_alpha}$ that O/Fe is tightly related to N/O ([O/Fe] $ \propto -1.2$ [N/O]; except at $z=4$, when there is no strong correlation), which is slightly less challenging to observe \citep[e.g.][]{AndrewsMartini2013,Masters2016}. Promisingly, the observational results from \cite{Steidel2016} are quantitatively consistent, see the data point in Fig. $\ref{fig:evolution_alpha}$. In EAGLE, at $z=0.1$, the gas-phase N/O ratio in star-forming galaxies decreases weakly as [N/O]$\propto -0.3$ log$_{10}$(sSFR) at fixed stellar mass (see the bottom-center panel in Fig. $\ref{fig:FMR_evo}$). We note that \cite{AndrewsMartini2013} find a potentially contradictory result (higher N/O ratio for galaxies with a higher SFR) at fixed mass M$_{\rm star} \lesssim 10^9$ M$_{\odot}$, but \cite{PerezMontero2013} find no secondary dependence of N/O on SFR at fixed mass. 

Stellar $\alpha$-enhancement may be estimated using absorption-line measurements, as has been done for massive elliptical galaxies \citep[e.g.][]{Trager2000,Conroy2013}. However, as the observed light from star-forming galaxies is dominated by hot massive stars, which typically have weaker metal absorption-lines, these measurements are more challenging for star-forming galaxies. An interesting approach could be to combine light-weighted stellar Fe/H abundances with gas-phase O/H abundances. This could be particularly useful at high sSFRs, where stellar and gas-phase $\alpha$-enhancements are predicted to be similar. In any case, the model predictions could be based on the same approach.

In summary, we have shown that the EAGLE simulation predicts the existence of a tight three-dimensional relation between mass, SFR and gas-phase $\alpha$-enhancement for galaxies with M$_{\rm star}=10^{9.0-10.5}$ M$_{\odot}$ due to a combination of rapid oxygen enrichment and delayed iron enrichment. A similar plane exists for stellar $\alpha$-enhancement as galaxies with higher SFRs have compressed SFHs at fixed mass \citep{MattheeSchaye2018b}. The gas-phase relation is tighter and, for $z<1$, evolves less than the ``fundamental metallicity relation'' between mass, SFR and metallicity. These results illustrate the promise of using chemical abundances to constrain galaxy evolution in a cosmological environment, but also imply that trends between physical properties inferred from observations may be affected by systematic variations in $\alpha$-enhancement.

\section*{Acknowledgments}
We thank the anonymous referee for their constructive comments. JM acknowledges the support of a Huygens PhD fellowship from Leiden University. We thank Jarle Brinchmann, Rob Crain and David Sobral for discussions. We acknowledge the use of the {\sc Topcat} software \citep{Topcat} for assisting in rapid exploration of multi-dimensional datasets and the use of {\sc Python} and its {\sc numpy}, {\sc matplotlib} and {\sc pandas} packages.




\bibliographystyle{mnras}

\bibliography{bibliography_pceagle.bib}





\bsp	
\label{lastpage}

\end{document}